\documentclass[conference]{IEEEtran}

\usepackage[utf8]{inputenc}
\usepackage[T1]{fontenc}
\usepackage[english]{babel}
\usepackage[final,babel]{microtype}
\usepackage[shortcuts]{extdash}
\usepackage[noadjust]{cite}

\usepackage{amsmath}
\usepackage{amssymb}
\usepackage{mathtools}
\usepackage{dsfont}
\allowdisplaybreaks

\usepackage{booktabs}
\usepackage{multirow}
\usepackage[position=top,font=small,caption=false]{subfig}
\usepackage{float}
\usepackage{xcolor}
\usepackage{url}
\usepackage{textcomp}

\usepackage{graphicx}
\graphicspath{{fig/}}
\usepackage{tikz}

\newcommand\Ex{\mathds{E}}

\newcommand\argmin{\operatorname*{arg\,min}}

\makeatletter
\def\@IEEEsectpunct{.\ \,}
\def\paragraph{\@startsection{paragraph}{4}{\z@}%
{0.5ex plus 0.5ex minus 0.5ex}%
{0ex}{\normalfont\normalsize\bfseries}}
\makeatother

\begin{document}

\title{Semantic Compression for Edge-Assisted Systems}

\author{
  \IEEEauthorblockN{Igor Burago, Marco Levorato, and Sameer Singh}
  \IEEEauthorblockA{%
    Department of Computer Science\\
    University of California, Irvine\\
    Email: \{iburago,~levorato,~sameer\}@uci.edu%
  }
}

\maketitle

\begin{abstract}
A novel semantic approach to data selection and compression is
presented for the dynamic adaptation of IoT data processing and
transmission within ``wireless islands'', where a set of sensing
devices (sensors) are interconnected through one-hop wireless links
to a computational resource via a local access point.
The core of the proposed technique is a cooperative framework where
local classifiers at the mobile nodes are dynamically crafted and
updated based on the current state of the observed system, the global
processing objective and the characteristics of the sensors and data
streams.
The edge processor plays a key role by establishing a link between
content and operations within the distributed system.
The local classifiers are designed to filter the data streams and
provide only the needed information to the global classifier at the
edge processor, thus minimizing bandwidth usage.
However, the better the accuracy of these local classifiers, the
larger the energy necessary to run them at the individual sensors.
A formulation of the optimization problem for the dynamic construction
of the classifiers under bandwidth and energy constraints is proposed
and demonstrated on a synthetic example.
\end{abstract}

\section{Introduction}

\begin{figure}[t]
  \centering\includegraphics[width=0.74\columnwidth]{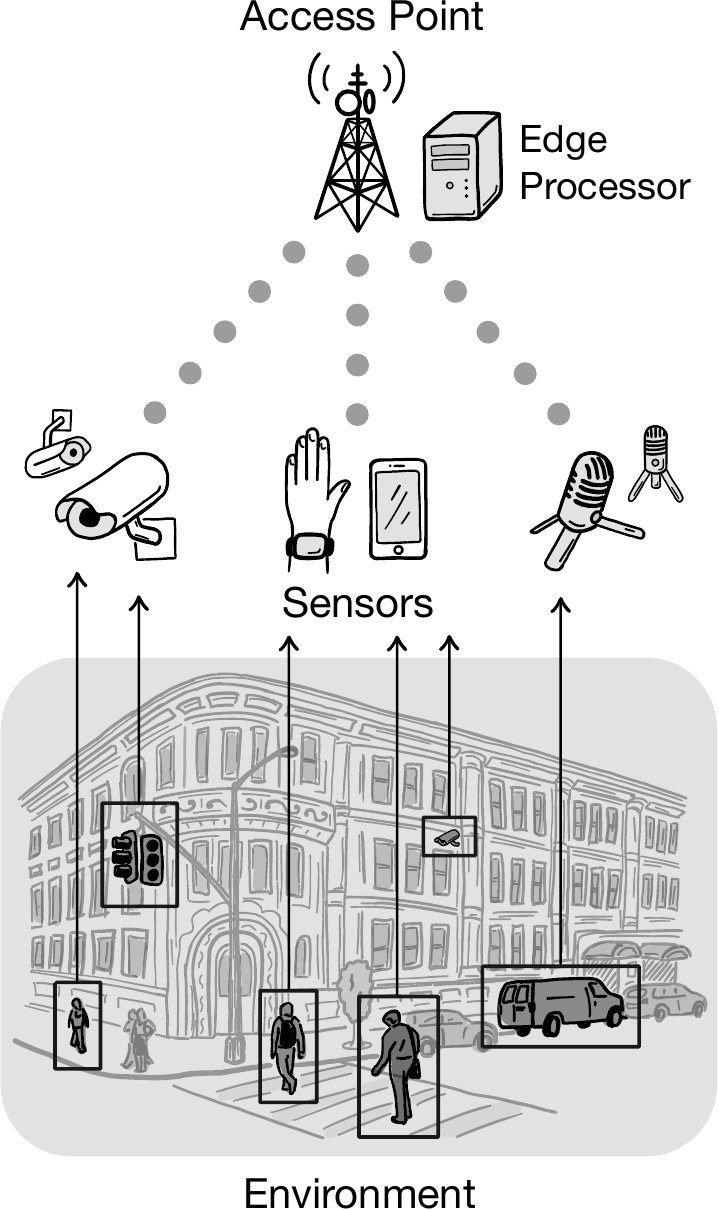}
  \caption{\textbf{Edge-assisted local network scenario:} A set of sensing devices acquire observations on the physical environment to support a global computational task performed at the edge processor.}
  \label{fig:system}
\end{figure}


The Internet of Things (IoT) paradigm~\cite{atzori2010internet} envisions a scenario where machines remotely interact to provide services and perform monitoring and control tasks.
To this aim, the IoT realizes a network of data sources, mobile devices, and processing centers interconnected through wireless and wireline links, where local and global algorithms cooperate in a distributed fashion.

Sophisticated large\-/scale application scenarios such as Smart City systems~\cite{neirotti2014current} and intelligent (or autonomous) vehicular networks~\cite{barba2012smart,martinez2010emergency} push the limits of IoT systems in sensing, communication and processing capabilities. To address the need for tight control loops, timely coordination and computation-intense processing, Fog and Edge Computing architectures~\cite{Bonomi,satyanarayanan2015edge} place computation resources at the edge of the wireless access infrastructure. In these architectures, mobile devices can offload computational tasks to edge data processors through one-hop low-latency links. The co\=/location of sensing and processing within a star topology allows reliable local coordination of remote devices informed by global resources, such as databases and data centers in the cloud.
However, the limited and time-varying bandwidth available in wireless environments makes the design of edge-based architectures challenging. This especially applies in those scenarios where IoT data streams coexist with other services on the same channel and network resource.

In this paper, we propose a framework for the dynamic adaptation of IoT data processing and transmission within ``wireless islands'', where a set of sensing devices (sensors) are interconnected with one-hop wireless links to a computational resource through a local access point (\emph{e.g.}, a cellular base station or a Wi\=/Fi access point). We specifically address an application scenario where the sensors and the edge processor cooperatively perform a real-time data acquisition and processing task, such as classification or detection based on environmental observations (see Fig.~\ref{fig:system}). The challenge, then, is to accomplish such task with the bandwidth, computational power and energy constraints imposed by the limited resources available at the device and network levels.

The core of the framework is a novel ``semantic'' approach to data selection and compression, where local classifiers at the mobile nodes are dynamically crafted and updated based on the current state of the observed system and its processing objective, together forming a continuously evolving \emph{context}.
The edge processor plays a key role by establishing a link between \emph{content} and \emph{operations} within the distributed system.
The local classifiers are designed to filter the data streams and provide only the needed information to the global classifier at the edge processor, thus minimizing bandwidth usage.
However, the better the accuracy of these local classifiers, the larger the energy necessary to run them at the individual sensors.
Our framework builds on recent results~\cite{ribeiro16:kdd,ribeiro16:model-agnostic}, where classifier simplifications are applied to the problem of explaining the outcome of black box machine learning algorithms.

An interesting connection can be made to the traditional multimedia compression techniques, where the components imperceivable by humans are removed.
Thus, distortion of the original signal is accepted in those regions that are not needed by the final application.
This research extends this principle to data consumed by machines for general computational purposes.
Additionally, we expand the traditional focus on bandwidth compression by itself with the notion of energy-awareness. 

The rest of the paper is organized as follows.
Section~\ref{sec:probform} introduces the general scenario and describes the problem addressed herein.
In Section~\ref{sec:semcompr}, we present the semantic compression framework, and illustrate its key components on an examplary problem in Section~\ref{sec:numres}.
Section~\ref{sec:concl} concludes the paper.

\section{Problem Formulation}
\label{sec:probform}

\begin{figure*}[t]
  \centering\includegraphics[width=0.94\textwidth]{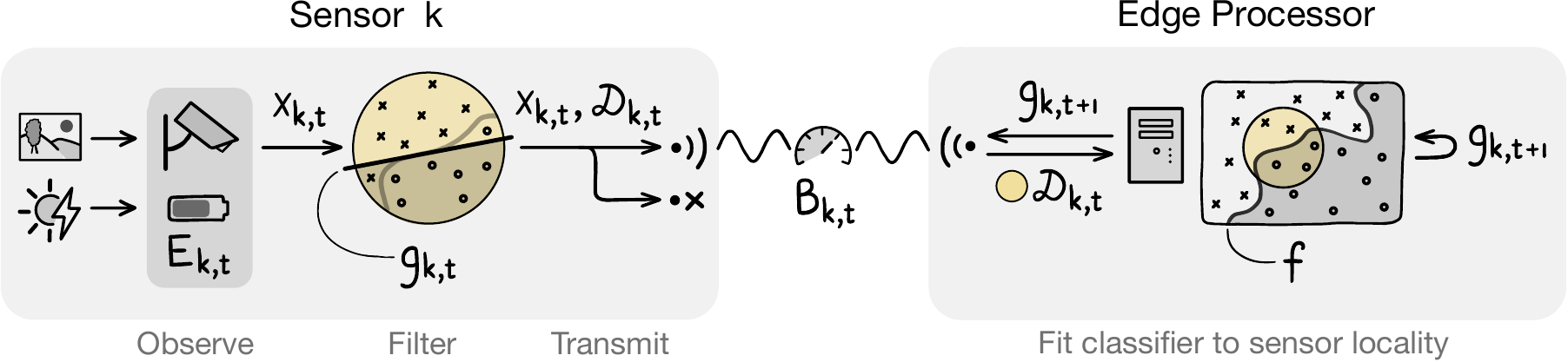}
  \caption{\textbf{Illustration of the problem:} Dynamic energy- and bandwidth-aware adaptation of local data filtering serving the purpose of global estimation. The figure illustrates the components of the acquisition, communication, processing and control for one sensor.}
  \label{fig:filtering}
\end{figure*}

Recent advances in machine learning resulted in sophisticated models, which provide incredibly capable detectors of interest to IoT applications, particularly for image and video processing.
Instead of working only for niche or synthetic settings, these classifiers are able to handle real-world input from a large variety of environments.
As a consequence, the resulting classifiers often tend to be too complex in structure, and can only reside on devices capable of handling computationally-intense tasks.
However, mobile sensors collecting the data for processing have only limited observational power, computational capabilities, and energy availability.
Hence, due to constraints in these resources, they often cannot support such complex classifiers.
Fog and Edge architectures offer a solution to this issue by introducing computational resources within the local wireless island.
However, bandwidth constraints, often imposed by other competing services, limit the data that can be transferred from the sensors to the computational resources.
In these circumstances, pre-filtering the data at the sensors becomes necessary to avoid delay, data loss, or undesirable disruption of other wireless services.

A sketch of the architecture at the center of our studies is in Fig.~\ref{fig:system}, where a set of sensors acquire observations in some dynamic environment.
The sensors are wirelessly interconnected through a local access point (\emph{e.g.}, base station) to an edge processor.
The edge processor is assigned a computational task (possibly changing in time), such as the identification of human activities in public parks or traffic dangers in autonomous vehicles' networks.
This task corresponds to one or more classifiers taking the data streams from the sensing devices as their inputs.
The goal of the global classifiers is to achieve an average accuracy $\alpha$, measured in terms of classification errors.

For the sake of explanation ease, we introduce the notion of temporal period, where time is discretized and indexed with~$t$.
The~$K$ sensors are connected to the edge processor through wireless links of capacity $b_{k,t}$, $k=1,\ldots,K$, in the period~$t$.
A constraint $b_t$ on the overall capacity available to the sensors, where $\sum_{k=1}^K b_{k,t}\leq b_t$, can be introduced to capture channel sharing.
The signal acquired by a sensor~$k$ in the time period~$t$ is $X_{k,t}$.
Each sensor has an energy storage for processing and transmission, where the amount of energy available at sensor~$k$ in period~$t$ is equal to $e_{k,t}$.
The energy storage can be refilled through charging or energy harvesting, modeled as a random arrival process.
The goal of the system is to guarantee the wanted accuracy at the edge processor using the available bandwidth and energy.
Fig.~\ref{fig:filtering} illustrates the components of the system for an individual sensor~$k$.

The sensors implement local classifiers which serve the purpose of filtering out \emph{unusable data}, defined as the data that are not needed for maintaining the target accuracy at the edge processor. While the amount of data transferred from the sensors to the edge is bounded by the time-varying capacity of the channel, the efficacy in locally removing unnecessary data is bounded by the processing power and energy availability at the sensors. On the one hand, the transmission of unfiltered data may violate the bandwidth constraint, thus causing data loss and disruption of existing wireless services. On the other hand, running a complex local classifier may require excessive computational effort and energy expense to the mobile devices.

We formulate an optimization problem capturing the tension between these two extremes for the purposes of dynamic adaptation of filters deployed at the sensors.
Based on the input from the sensors, the edge processor periodically produces a new filter with controlled complexity for each sensor, based on bandwidth and energy usage constraints following from high-level operational objectives.
Herein, we focus on building customized classifiers possessing the following characteristics:

\begin{itemize}
\item \textbf{Locality.} The sensor-specific classifiers will be trained to achieve a certain accuracy level for the kinds of inputs the sensor is likely to receive. For instance, the local classifiers will be built to provide low-error predictions for indoor images if the sensor is placed inside.
\item \textbf{Bandwidth-Awareness.} The local classifiers are designed to be used as bandwidth-preserving filters, thus optimizing for the false-negative rate to meet the bandwidth constraints imposed by the link to the global edge processor.
\item \textbf{Complexity and Energy-Awareness.} The design of the local classifiers will satisfy complexity and energy requirements of the sensor as determined by a stochastic energy-arrival process.
\end{itemize}

Given the complex, accurate classifier at the edge, our objective is to build a sensor-specific classifier tailored to the distribution of samples in the current period, and satisfying the bounded complexity and bandwidth usage.
More formally, we are provided with a pre-trained binary classifier, \emph{e.g.}, one detecting whether a person is visible by the sensor, denoted by $f\colon X\to\{0,1\}$, where $X$ is the space of possible inputs.
We treat this classifier as a black-box function in order to support as wide of a variety of machine learning algorithms as possible.

For a sensor~$k$ during period~$t$, the goal is to identify a \emph{local classifier} $g_{k,t}\colon X\to\{0,1\}$, $g_{k,t}\in\mathcal{G}$, that meets the specifications of the sensor, where $\mathcal{G}$ is the family of machine learning classifiers we want the sensor to use (for instance, linear classifiers).
In particular, we are provided with the following requirements corresponding to the aforementioned characteristics:
\begin{itemize}
\item \textbf{Locality $\mathcal{D}_{k,t}$:} The expected distribution of the sensor inputs for period~$t$ is denoted by $\mathcal{D}_{k,t}$. We want $g_{k,t}$ to be as accurate as $f$ as possible on inputs from this distribution.
\item \textbf{Bandwidth $b_{k,t}$:} The average amount of data \emph{allowed} to be transmitted by $g_{k,t}$ for the period $t$ should be less than $b_{k,t}$. (It is also possible to consider a generalization where only the total capacity $b_t$ for all sensors is provided.)
\item \textbf{Energy $e_{k,t}$:} Average energy used by $g_{k,t}$ for the period $t$ should be less than $e_{k,t}$.
\end{itemize}
In this work we assume that the customized classifier $g_{k,t}$ will be built on the edge, not the sensor, and thus the computational efficiency of estimating $g_{k,t}$ is not restricted.

\section{Semantic Compression}
\label{sec:semcompr}

In this section, we outline our proposed approach to constructing a classifier $g_{k,t}$ that meets the sensor's requirements on energy, bandwidth, and locality for the period~$t$, while still being faithful to the complex, global classifier~$f$.

\paragraph*{Energy Efficiency}
The primary obstacle with using~$f$ at the sensor level is its computational complexity.
For insance, each prediction by a neural network can often take hundreds to thousands of floating-point computations, resulting in a heavy power consumption.
Instead, we are concerned with learning an \emph{energy-efficient} classifier $g_{k,t}\in\mathcal{G}$, for $\mathcal{G}$ being limited to a simpler model family, such as SVMs, decision trees, linear classifiers, \emph{etc}.
We define the energy consumed by $g_{k,t}$ for an input as $E_{g_{k,t}}\colon X\to\mathbb{R}_{\ge0}$; the average energy used by the sensor~$k$ for period~$t$ will be $\Ex_{x\sim\mathcal{D}_{k,t}}[E_{g_{k,t}}(x)]$.
We also define a penalty on the classifier for violating an energy constraint $e_{k,t}$ as $R_E$, such that $R_E(E_{g_{k,t}}(x),e_{k,t})=0$ if $g_{k,t}$ meets the energy requirement $e_{k,t}$, and $R_E(E_{g_{k,t}}(x),e_{k,t})>0$ otherwise.
Since directly estimating the energy consumption $E_{g_{k,t}}$ of a classifier $g_{k,t}$ is challenging, we use the number of computational operations as a proxy, and thus $R_E$ penalizes $g_{k,t}$ the more operations it requires for a prediction.

\begin{figure}[t]
  \centering\includegraphics[width=0.74\columnwidth]{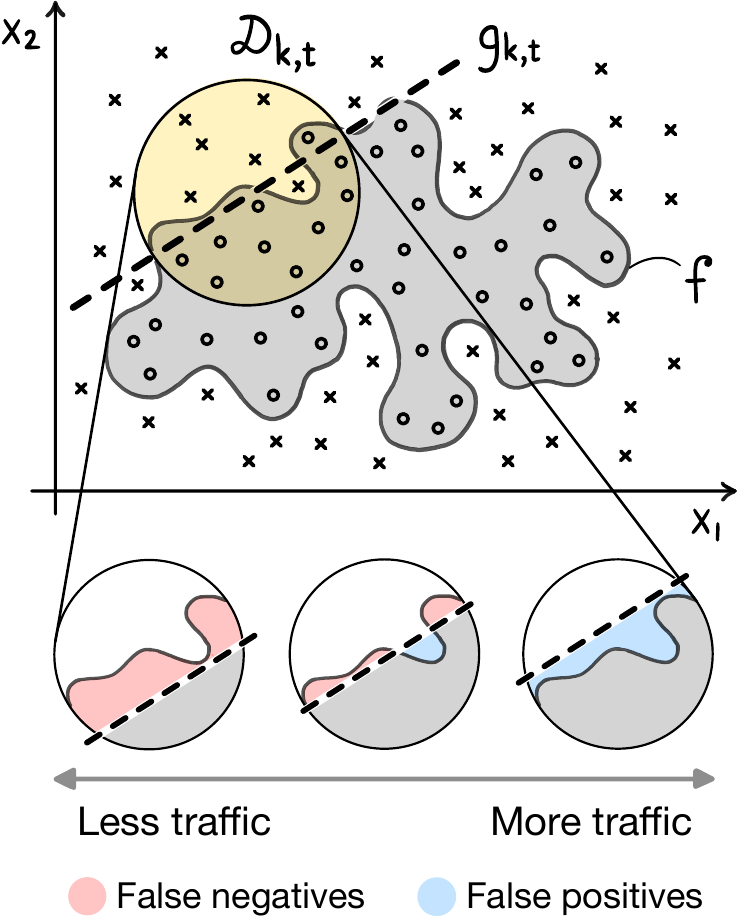}
  \caption{\textbf{Localized semantic classifier compression:}
  The gray area depicts the subspace of positive detections of a global black-box classifier~$f$ in the space of all inputs~$X$. The dashed bold line represents a simplified linear classifier $g_{k,t}$ chosen to fit~$f$ only in the locality $\mathcal{D}_{k,t}$ of recent inputs from a sensor~$k$ (yellow circle), and so does not need to bear the full complexity of~$f$.
  Our approach draws instances from $\mathcal{D}_{k,t}$, classifies them with~$f$, and uses the resulting sample for optimizing $g_{k,t}$. Due to energy and bandwidth constraints, different boundaries $g_{k,t}$ may be obtained as illustrated under the plot: Aggressive ones save traffic by capturing less but risk frequent misses (left example); conservative ones avoid misses by capturing more but generate more traffic (right example).}
  \label{fig:localization}
\end{figure}

\paragraph*{Locality}
Obviously, an energy-efficient classifier $g_{k,t}$, by using a simpler structure, cannot have the same general representation capabilities as the global classifier~$f$ for the complete range of inputs.
However, in any given time period, most sensors do not receive the full variety of inputs that the global classifier is designed to support, and thus it is possible to have $g_{k,t}$ focus its representation on the inputs expected at the sensor.
In order to identify such a $g_{k,t}$, we use the expected distribution of inputs, $\mathcal{D}_{k,t}$, to compute how similar $g_{k,t}$ is to~$f$.
In particular, given a \emph{loss function} $L(f(x),g_{k,t}(x))$ between $g_{k,t}$'s and $f$'s predictions on an instance $x$, \emph{e.g.}, the squared loss $L_{\text{sq}}(a,b) {\,=\,} (a-b)^2$ or the logistic loss $L_{\text{ll}}(a,b) = -a\log b - (1-a)\log(1-b)$, we evaluate the similarity between $g_{k,t}$ and $f$ as $\Ex_{x\sim\mathcal{D}_{k,t}}\big[L(f(x), g_{k,t}(x))\big]$.
Fig.~\ref{fig:localization} illustrates the intuition, where a complex, power-consuming global classifier~$f$ (solid gray curve) can be approximated quite well \emph{locally} by a simple, and thus energy-efficient, classifier $g_{k,t}$ (dashed bold line).

\paragraph*{Bandwidth Awareness}
Every automated detector is accompanied by a certain level of expected error, often measured as the rate of false positives and false negatives.
Due to the energy constraints on the desired classifier $g_{k,t}$, it may not be able to maintain the same low error levels as the global classifier~$f$, even on the local distribution of inputs.
In such situations, we can treat $g_{k,t}$ as the sensor-level filtering of the inputs, with $f$ running at the edge level to achieve the same low error levels.
Thus there is a trade-off between how much the bandwidth is used to transmit false positives versus missing out a relevant input in order to conserve the bandwidth.
We define the amount of data $g_{k,t}$ will use for an input~$x$ as $B_{g_{k,t}}\colon X\to\mathbb{R}_{\geq0}$; the average data transmitted by the sensor for period $t$ will be $\Ex_{x\sim\mathcal{D}_{k,t}}[B_{g_{k,t}}(x)]$.
We further define the penalty on the classifier $g_{k,t}$ for violating the bandwidth $b_{k,t}$ as $R_B$, such that $R_B(B_{g_{k,t}}(x),b_{k,t})=0$ if $g_{k,t}$ uses less than $b_{k,t}$ bandwidth, and $R_B(B_{g_{k,t}}(x),b_{k,t})>0$ otherwise.
%
%
Fig.~\ref{fig:localization} shows an example where a classifier that is not aware of its use as a filter (the leftmost example) may transmit less but have a high error rate, while a bandwidth-aware classifier (in the middle) will obtain lower false negative rate.

\paragraph*{Semantic Compression}
From the sensor specifications, namely local distribution $\mathcal{D}_{k,t}$, energy consumption constraint $e_{k,t}$ and penalty function $R_E$, bandwidth constraint $b_{k,t}$ and penalty function $R_B$, and the global classifier $f$, we can frame the search for the sensor-specific classifier $g_{k,t}$ as the following optimization problem to be solved periodically over time:
\begin{align} \label{eq:sem:local}
  g_{k,t}^* &= \argmin_{g_{k,t}\in\mathcal{G}}\, \Ex_{x\sim\mathcal{D}_{k,t}}\big[L(f(x),g_{k,t}(x))\big],\\
  \text{s.t. } \label{eq:sem:local:constr-energy}
  &\Ex_{x\sim\mathcal{D}_{k,t}}\big[ R_E\big(E_{g_{k,t}}(x), e_{k,t}\big) \big] \leq\varepsilon_{k,t},\\ \label{eq:sem:local:constr-bandw}
  &\Ex_{x\sim\mathcal{D}_{k,t}}\big[ R_B\big(B_{k,t}(x), b_{k,t}\big) \big] \leq\beta_{k,t}.
\end{align}
Here $\varepsilon_{k,t}$ and $\beta_{k,t}$ have the meaning of the tolerances on the expected penalties $R_E$ and $R_B$ for random observations following a given locality distribution $\mathcal{D}_{k,t}$.

The distribution $\mathcal{D}_{k,t}$ serves a proxy role conveying to the edge processor a local description of expected observations at the sensor, without wasting the bandwidth for transmitting the observations themselves.
The edge processor, in turn, replies to the sensor with a classifier $g_{k,t}^*$, locally tuned to $\mathcal{D}_{k,t}$ according to the problem in Eq.~\eqref{eq:sem:local}--\eqref{eq:sem:local:constr-bandw}.
For each particular sensor and time period, the distribution $\mathcal{D}_{k,t}$ is fixed, so the efficacy of this semantic compression scheme is determined by whether the family of local classifiers~$\mathcal{G}$ is flexible enough for the distribution of positive and negative samples in $\mathcal{D}_{k,t}$.
However, at a larger scope, the locality $\mathcal{D}_{k,t}$ may vary and is subject to negotiation between the sensor and the edge processor.

With the shape of the locality $\mathcal{D}_{k,t}$ controllable, the quality of corresponding classifiers $g_{k,t}$ may be improved additionally through locality tuning.
This brings the option to view the optimization in Eq.~\eqref{eq:sem:local}--\eqref{eq:sem:local:constr-bandw} as a subproblem for a higher-level control task, maintaining a desired aptitude of classifiers $g_{k,t}$ on a sequence of observations generated by the sensor.
This way, the problem of finding optimal local $g_{k,t}$ may be extended to the broader adaptive-control problem of maintaining a desired accuracy of filtration by adjusting the locality-capturing procedure delivering distributions $\mathcal{D}_{k,t}$, such that
\begin{equation} \label{eq:sem:control}
  \Ex_{x\sim\mathcal{S}_{k,t}}\big[ Q_{g_{k,t}}(x,\mathcal{D}_{k,t}) \big] \le q_{k,t}.
\end{equation}
The penalty function $Q_{g_{k,t}}$ stands for the losses we bear from any inadequacies of the local classifier $g_{k,t}^*$ to the particular choice of locality $\mathcal{D}_{k,t}$, which we would like to keep bounded by a tolerance $q_{k,t}$.
Here, the quality is monitored for inputs from some control distribution $\mathcal{S}_{k,t}$ chosen by the edge processor using the empirical data arriving from the sensor and the a~priori strategic objectives for the ultimate outcomes of the sensor-edge system as a whole.
In practice, $\mathcal{S}_{k,t}$ may coincide with the global observatory distribution~$\mathcal{X}$, the locality distribution $\mathcal{D}_{k,t}$, or can be derived from the sequence of empirical observations obtained by the sensor~$k$.
In Eq.~\eqref{eq:sem:control}, the locality $\mathcal{D}_{k,t}$ is made an argument of the penalty $Q_{g_{k,t}}$ to highlight its potential role as the control ``variable''.
One simple example giving an idea of how localities $\mathcal{D}_{k,t}$ may be parametrized and controlled will be given in the following section.

\section{Simulation Results}
\label{sec:numres}

In order to illustrate the feasibility of the proposed approach, let us
consider a motivating example of a binary classification problem,
in the context of a single sensor-edge pair (for this reason we omit
the index~$k$ below, for the sake of brevity).

As customary, input observations subject to classification come as
feature vectors in a multidimensional vector space~$X$. The two
classes correspond to the sets of observations that are to be registered
by the sensor-edge system (positives), versus the rest (negatives).
In this case, probability distributions of both classes are set to
be Gaussian mixtures (and so is, therefore, the joint distribution~$\mathcal{X}$).
Both mixtures consist of the same number of symmetric normally-distributed
components centered equidistantly on a number of lines parallel to
the main diagonal of the unit hypercube.


For simplicity, we assume that both $f$ and $g\in\mathcal{G}$ belong
to the same class of Support Vector Machine classifiers~(SVMs) working
in the space~$X$. To satisfy the requirement of $g$ having a lower
complexity than~$f$, the class~$\mathcal{G}$ is limited to SVMs
with linear kernels, while the reference global classifier~$f$ is
trained for the kernel of Gaussian radial basis functions (and can
be replaced with even more computationally-intense classifier).
Each locality distribution~$\mathcal{D}_t$ guiding the selection
of training samples for the on-sensor classifiers $g_t$ is set to
be a uniform distribution in a sphere described by its center and
radius~$r_t$. By the nature of the distribution~$\mathcal{X}$,
the local and global accuracy of the classifier~$f$ is expected
to not differ significantly, while the accuracy of localized classifiers~$g$
shall be sensitive to the localities~$\mathcal{D}_t$ and their
sizes~$r_t$.

In this circumstances, the applicability of the problem statements
given in Section~\ref{sec:probform} to this detection task requires
a study of two aspects of the system: (i)~The accuracy of the localized
classifiers~$g_t$ for different spheres~$\mathcal{D}_t$ as
a function of radii~$r_t$ and the update frequency~$1/\gamma$.
(ii)~Realization of actual distributions of consecutive observations~$x_t$
in the data for a desired update frequency, and the procedure
for adaptively choosing the radii~$r_t$ reacting to the accuracy-complexity
tradeoff.

To this end, both in this specific example and in general, we need
to be in possession of two samples. First, a labeled training dataset
of pairs~$(z_i,y_i)$ is necessary, where points $z_i\in X$ are drawn
from the joint distribution of observations~$\mathcal{X}$,
and $y_i\in\{0,1\}$ signify the corresponding labels.
We can assume the availability of this sample~$Z$ without any loss
of generality, as the very problem setting given in Section~\ref{sec:probform}
starts with a classifier~$f$ that has to be trained on some sample,
which we can reuse here for~$Z$. In the unsupervised case, for the
purposes of the following discussion, labels~$y_i$ can be defined
by the outcomes~$f(z_i)$ of the global classifier~$f$.

Second, it is necessary to have a sample of one or more trajectories
$S{\,=\,}(x_1,\ldots,x_T)$, $x_t{\,\in\,}X$ representative of the
sequential process generating observations on the sensor. In practice,
this sample can be obtained from previous, nonadaptive runs of the
sensor-edge system in question, where all sensor observations
eventually reach and get accumulated at the edge processor.
In this example problem, we assume that the trajectory
distribution~$\mathcal{S}$ follows the general
distribution~$\mathcal{X}$ (which would also likely be the case in
general, as well, unless the nature of observation process dictates
otherwise). Adhering to this assumption, we generate a sample~$S$
as a Markov chain starting from a randomly chosen
point~$x_0\sim\mathcal{X}$ and continuing by applying
the Metropolis-Hastings sampling algorithm to
the distribution~$\mathcal{X}$.

\begin{figure}[t]
  \centering
  \includegraphics[width=\columnwidth]{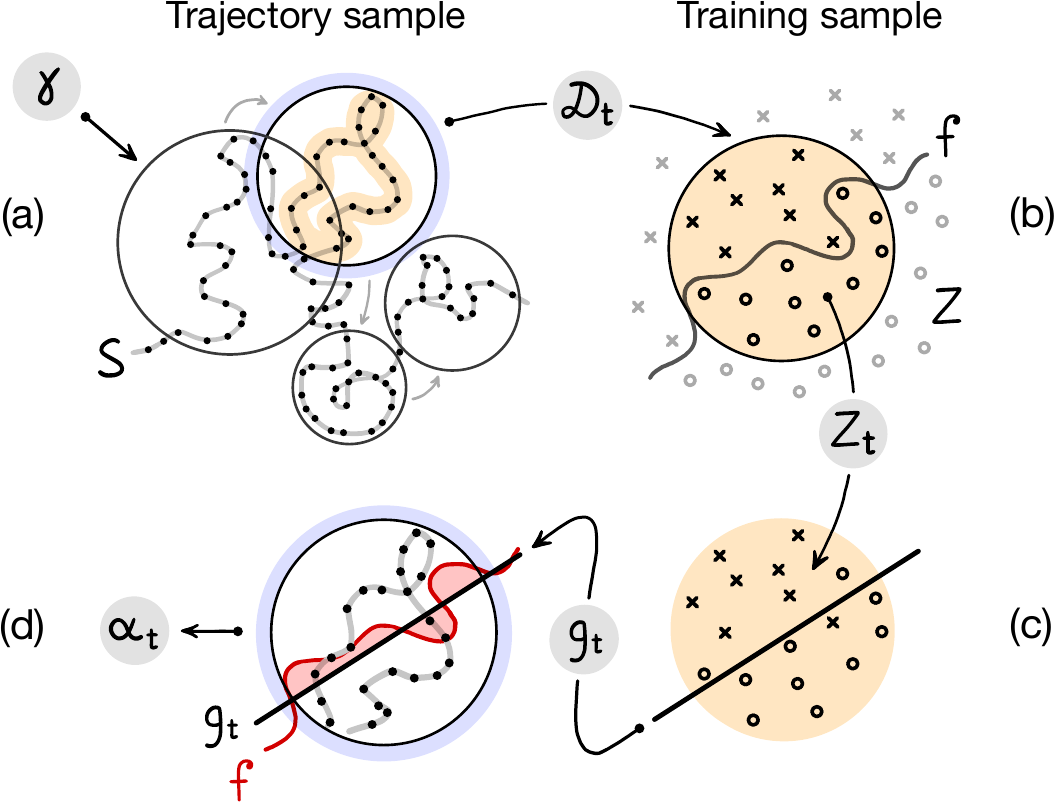}
  \caption{\textbf{Trajectory sampling procedure:}
  Schematic representation of the stages (a)--(d),
  highlighting the key variables invloved.}
  \label{fig:trajectory-sampling}
\end{figure}

The two aforementioned aspects of the system, then, can be studied
through the following duplex sampling procedure (schematically
depicted in Fig.~\ref{fig:trajectory-sampling}).
\begin{enumerate}
\item For each update frequency~$1/\gamma$ (or, equivalently, the length
of the update period~$\gamma$ in the number of observations), draw
a sample of subsequences $S_t(\gamma)=(x_{t-\gamma+1},\ldots,x_t)$
of~$\gamma$ consecutive observations along the trajectory~$S$.
\item For each subsequence~$S_t(\gamma)$:
\begin{enumerate}
\item Find the minimal sphere~$\mathcal{D}_t$ containing all (or a given
percentage of) points $x_{t-\gamma+1},\ldots,x_t$.
\item Sample points $Z_t(\gamma)=\{(z_i,y_i)\in Z\mid z_i\sim\mathcal{D}_t\}$
from the general training sample~$Z$ uniformly inside of the sphere~$\mathcal{D}_t$.
\item Using the points in $Z_t(\gamma)$ as a training sample, fit a classifier~$g_t\in\mathcal{G}$
to a desired quality.
\item Apply the classifier~$g_t$ to the points in the subsequence~$S_t(\gamma)$,
comparing the verdicts of~$g_t$ to the corresponding verdicts
of the reference classifier~$f$ for those same points in~$S_t(\gamma)$.
\item Store the radius~$r_t(\gamma)$ of the sphere~$\mathcal{D}_t$
and the resulting accuracy~$\alpha_t(\gamma)$ of the localized
classifier~$g_t$ on the points in~$S_t(\gamma)$.
\end{enumerate}
\end{enumerate}

With the accumulated statistics of radii~$r_t(\gamma)$ and
accuracies~$\alpha_t(\gamma)$, it is then possible for us to compute
the empirical averages of both of these features over trajectory's
subsequences as functions of update period~$\gamma$.

\begin{figure}[t]
  \centering\includegraphics{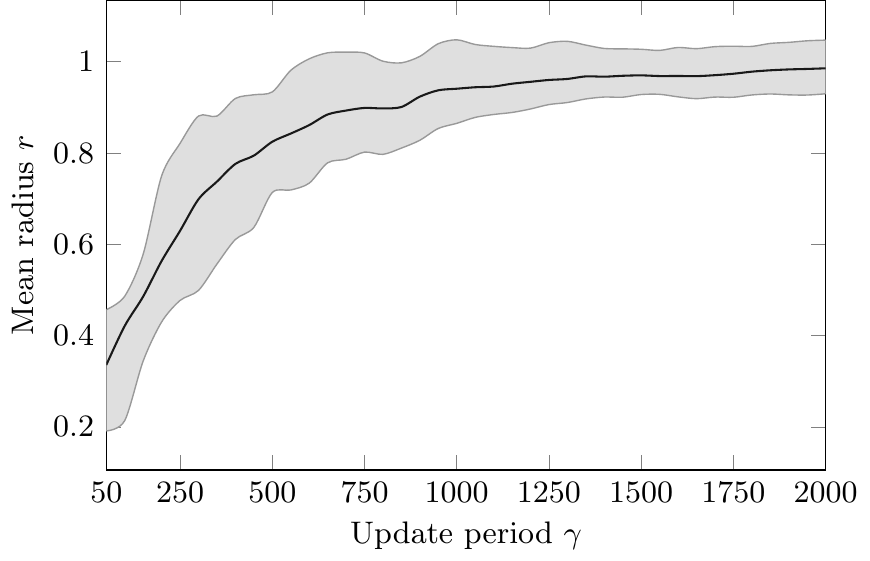}
  \caption{\textbf{Update locality:}
  Average radius of a sphere containing $95\%$ of points
  in trajectory subsequences, as a function of update period for
  the example problem. The range between 0.25- and 0.75-quantiles
  is highlighed in gray.}
  \label{fig:synth-radius}
\end{figure}

\begin{figure}[t]
  \centering\includegraphics{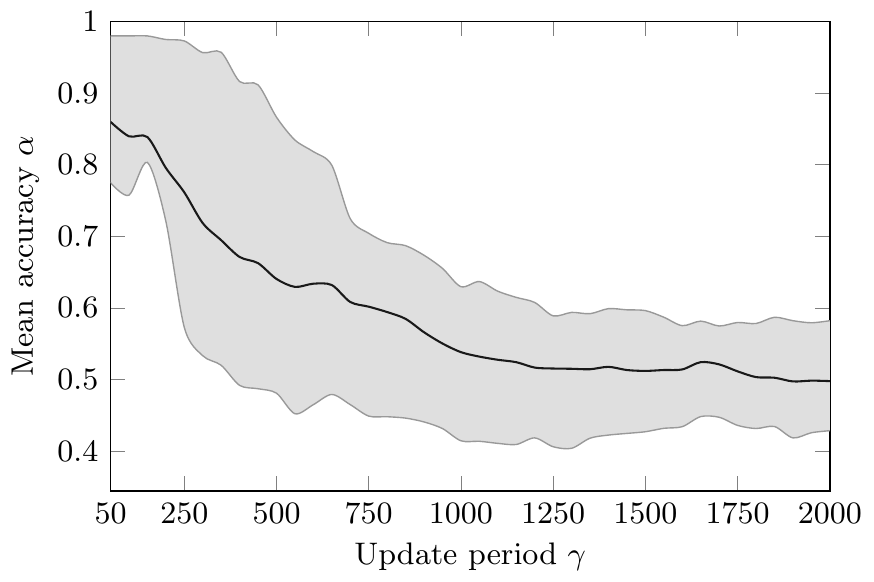}
  \caption{\textbf{Local accuracy:}
  Average accuracy as a function of update period for
  the example problem. The range between 0.25- and 0.75-quantiles
  is highlighed in gray.}
  \label{fig:synth-accuracy}
\end{figure}

Fig.~\ref{fig:synth-radius} and~\ref{fig:synth-accuracy} demonstrate
these functional relations in the case of our motivating example for
a multidimensional Gaussian sample.
The former figure depicts the average radius of spheres containing
$95\%$ of the points in subsequences~$S_t(\gamma)$ for different
values of $\gamma$.
As we can see, the average radius quickly grows as the update period
increases.
The latter figure highlights the opposite trend: the accuracy of
locally-fit classifiers~$g_t$ almost monotonically decreases with
increasing period of updates.
For comparison, the accuracy of the global classifier~$f$ when it is
implemented as an RBF-kernel SVM fluctuates insignificantly around
$98\%$ independent of the update frequency~$\gamma$.

\begin{figure}[t]
  \centering\includegraphics{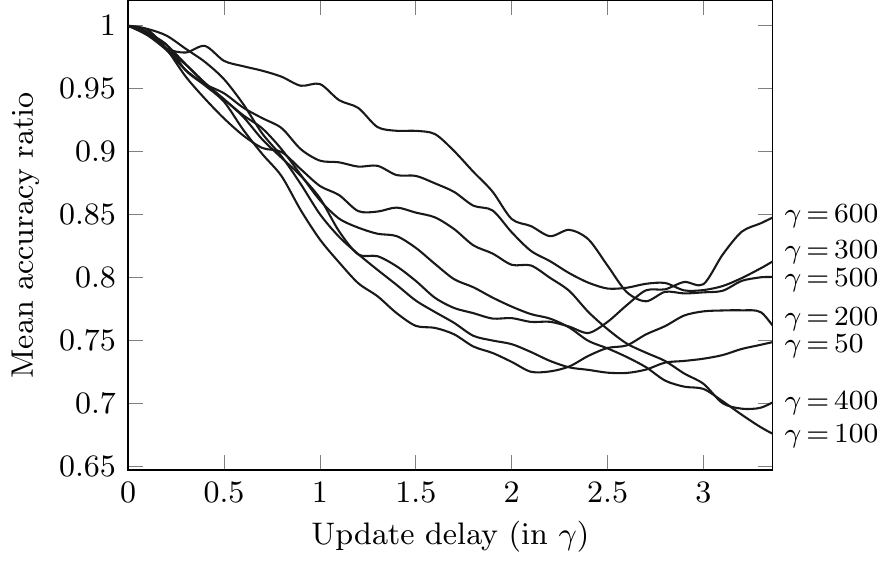}
  \caption{\textbf{Local classifier aging:}
  Relative change in average accuracy of a local classifier
  $g_t$ for different update periods $\gamma$ as a function
  of update delay normalized by the length of update period.}
  \label{fig:synth-accuracy-aging}
\end{figure}

Here both~$f$ and~$g_t$ were trained to treat both false positive
and false negatives equally; in cases where it is intolerable to miss
detections due to localized approximation, the same trends will be
present for respectively adjusted~$g_t$.
The choice of update frequency can be guided by the penalty taken
by the accuracy~$\alpha_t(\gamma)$ when the classifier~$g_t$ trained
for a locality~$\mathcal{D}_t$ is kept for a use in the subsequent
localities $\mathcal{D}_{t+1},\mathcal{D}_{t+2},\ldots\,$ without
an update.
For our example this relation is summarized in
Fig.~\ref{fig:synth-accuracy-aging} showing the change in mean
accuracy of a local classifier~$g_t$ as a function of the delay
between its training and its usage.
The x-axis measures the delay relative to the update period length
$\gamma$.
The y-axis measures the ratio between the mean accuracy for
the trajectory subsequence corresponding to the moment a local
classifier~$g_t$ is used and the mean accuracy for the trajectory
subseqence corresponding to the moment locality $\mathcal{D}_t$
was captured.

All three of these relations confirm the feasibility of the
assumptions underlying the problem formulation, that, while simpler
local classifiers~$g_t$ have poor accuracy globally, their quality
catches up for frequent locality updates to a satisfactory level
comparable to that of the global classifier~$f$.
The ultimate quality of the resulting system will, of course, depend
significantly on the mutual compatibility of the data
distribution~$\mathcal{X}$ (governing the complexity of the global
classifier~$f$), the family of local classifiers~$\mathcal{G}$, the
form of locality distributions~$\mathcal{D}_t$, and the constraints on
the desired accuracy.
For instance, when sensor sampling trajectories do not exhibit enough
compactness as measured by the form~$\mathcal{D}_t$ and~$g_t$,
it might be problematic or even impossible to achieve very high levels
of accuracy with the localized substitution classifiers~$g_t$.
In each particular case, the limits of the achievable results should
be studied separately, e.g., using the above trajectory-sampling
procedure.

For problems where, like in our example here, the locality of the
space~$X$ can be exploited well for a given $\mathcal{D}_t$ and $g_t$,
it opens the possibility for an efficient adaptation of the
locality~$\mathcal{D}_t(\tau)$ as a function of some control
parameters~$\tau$.
For instance, here the update period~$\gamma$ can serve the role of
the parameter~$\tau$, with the control objective consisting in keeping
it smaller than some~$\gamma_{0}$ guaranteeing a desired accuracy
(according to Fig.~\ref{fig:synth-accuracy}).

\section{Conclusions}
\label{sec:concl}

Sophisticated IoT systems often involve combining sensing, communication, and processing capabilities.
Recent architectures for such IoT systems often perform expensive computation at the edge-level, in order for the mobile devices to utilize their limited energy for sensing and transmission.
However, such architectures often cannot meet the tight constraints of a time-varying or limited bandwidth availability, as is common in real world applications, due to their need to communicate all of the data from the sensor-level devices to the edge.

In this paper, we proposed an alternative architecture where the edge and the devices perform the computation cooperatively.
The core of our proposed approach is to provide a ``semantic'' strategy for carrying out this sharing of the computation: we dynamically craft customized classifiers for each sensor that define what the sensor device will communicate to the edge processor, thus offloading majority of the computation to these devices.
This proposed design of sensor-specific classifiers takes into account the various properties of the current context such as the sensor-specific distribution of inputs that the device is likely to observe, the energy resources and constraints on the device, and the time-varying limitations on the shared bandwidth to the edge.

We showed the feasibility of our semantic approach using simulated experiments.
We demonstrated that simple, energy-efficient classifiers can be as accurate in classification as complex classifiers if we utilize the distribution inputs that the sensing device is likely to receive when constructing them.
We further showed that the approach is fairly robust to changes in this distribution of inputs over time. 
Although the classifiers need to be updated as the current context of the sensors and the edge changes over time, we also demonstrated that the sensor-specific classifiers still maintain accuracy even if they are not updated very frequently.
With these encouraging results, we are interested in future to deploy such an architecture to real-world IoT testbeds.

\bibliographystyle{IEEEtran}
\bibliography{IoT,problem,sameer}

\end{document}